\documentclass{aip-cp}

\usepackage[numbers]{natbib}
\usepackage{rotating}
\usepackage{graphicx}

\begin{document}

\title{Quantum Instabilities of Solitons}

\author[aff1]{H. Weigel\corref{cor1}}

\affil[aff1]{Institute for Theoretical Physics, Physics Department,
Stellenbosch University, Matieland 7602, South Africa}

\corresp[cor1]{Corresponding author: weigel@sun.ac.za}
\maketitle

\begin{abstract}
We compute the vacuum polarization energies for a couple of soliton models in one space 
and one time dimensions. These solitons are mappings that connect different degenerate vacua. 
From the considered sample solitons we conjecture that the vacuum polarization contribution
to the total energy leads to instabilities whenever degenerate vacua with different curvatures 
in field space are accessible to the soliton.
\end{abstract}

\section{INTRODUCTION}
Classical soliton solutions emerge frequently in $D=1+1$ theories with degenerate vacuum 
configurations. The prime examples are the kink within the $\phi^4$ model and the sine-Gordon 
soliton \cite{Ra82}. In these models the vacua are not only degenerate but also have identical 
curvatures in field space. Probably even more interesting are theories whose degenerate
vacua have different curvatures. To be specific, we call these different vacua primary and 
secondary, with the primary being approached by the soliton at (positive) spatial infinity.
Nevertheless, the soliton may dwell in the secondary vacuum within an arbitrarily large region 
of space. Typically the classical energy does not depend on the size of this region since the 
energy density is local and is non-vanishing only in regions in which the soliton configuration 
deviates from the degenerate vacua. On the contrary, quantum corrections are not local and thus 
they vary with the size of these regions. This is even more the case because the different 
curvatures translate into different masses of the quantum fluctuations. As examples we will 
consider the one-loop quantum correction, also known as the vacuum polarization energy (VPE),
for two soliton models with primary and secondary vacua.

We calculate the VPE from spectral methods that utilize scattering data \cite{Graham:2009zz}. 
Essentially the VPE is written as the change of the density of modes fluctuating around the 
soliton. According to the Krein formula \cite{Faulkner:1977aa} this change is related to the 
phase shift of the scattering problem with the potential generated by the soliton. Spectral 
methods make exhaustive use of the analytical properties of scattering data and eventually 
formulate the VPE as an integral of the Jost function for imaginary momenta. For the cases 
considered here, the formalism has been established in Refs. \cite{Weigel:2016zbs,Weigel:2017kgy}. 
The results from its numerical simulation agree with those from the heat kernel formalism, which, 
however, is significantly more intricate, {\it cf.} 
Refs. \cite{AlonsoIzquierdo:2011dy,Alonso-Izquierdo:2013pma}.

To streamline the presentation we will utilize dimensionless fields, coordinates and coupling 
constants. This produces an overall factor, call it $\lambda^{-1}$, for the Lagrangian. Then 
the coupling constant $\lambda$ does not appear in the field equations but upon quantization 
it produces a relative factor between the velocities of the fields and their conjugate momenta. 
Thus the classical energy is proportional to $\lambda^{-1}$ while the VPE does not scale
with $\lambda$.

\section{THE $\varphi^6$ MODEL}
\begin{figure}[ht]
  \centerline{\includegraphics[width=6.5cm,height=4cm]{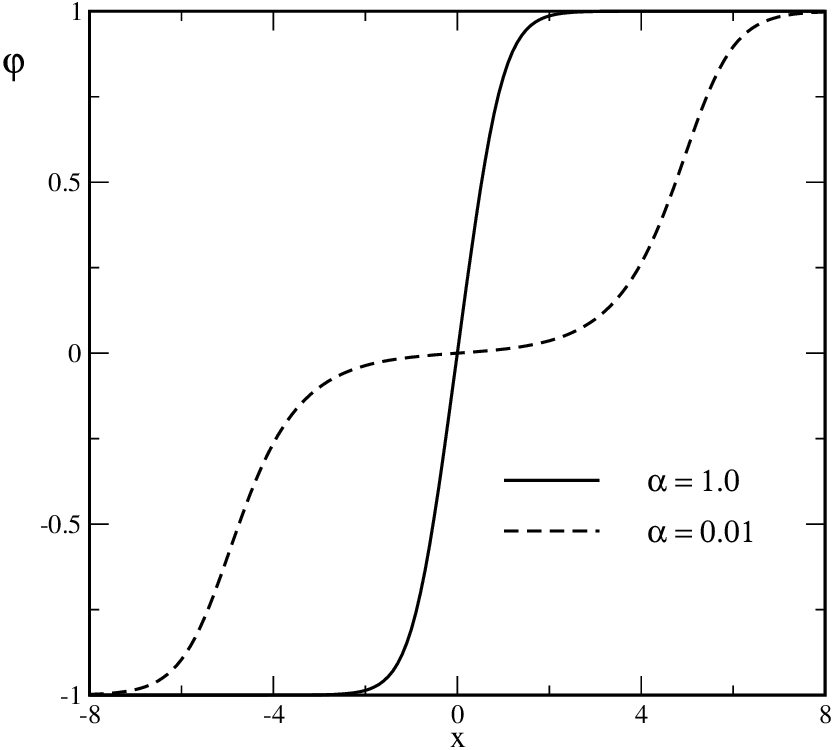}\hspace{1.9cm}
              \includegraphics[width=6.5cm,height=4cm]{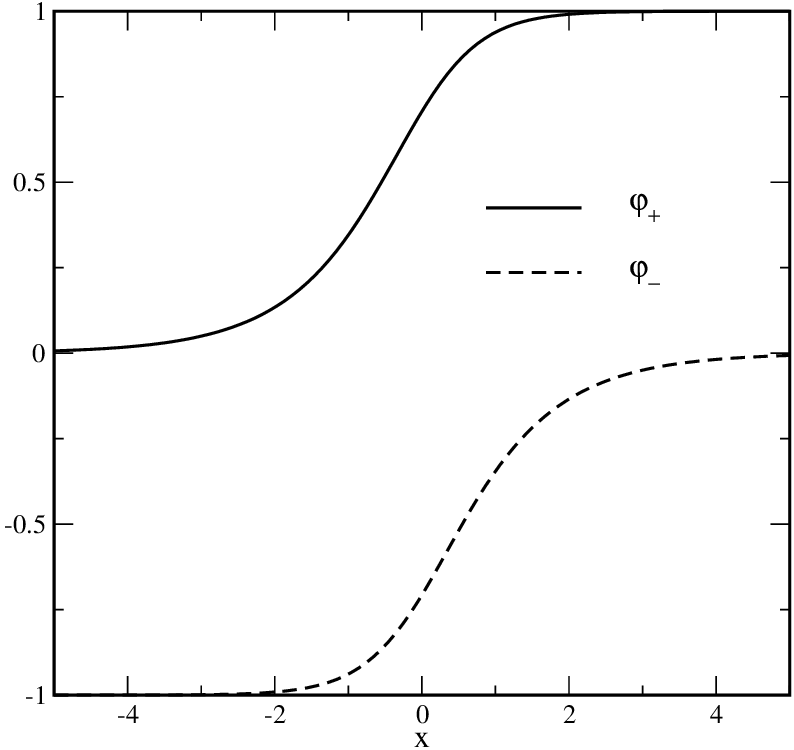}~}
  \caption{\label{fig:sol6}The soliton profiles of the $\varphi^6$ model. Note that 
$\alpha=0$ in the right panel.} 
\end{figure}

The scaled Lagrangian for the real scalar field $\varphi$ reads
\begin{equation}
\lambda \mathcal{L}=\frac{1}{2}\partial_\nu\varphi\partial^\nu\varphi-U(\varphi)
\qquad {\rm with} \qquad 
U(\varphi)=\frac{1}{2}\left(\varphi^2-1\right)^2\left(\varphi^2+\alpha^2\right)\,.
\label{eq:phi6}\end{equation}
For non-zero values of the model parameter $\alpha$ only primary vacua at $\varphi=\pm1$ 
exist. Solitons exist that mediate between these two configurations as the spatial coordinate 
varies between negative and positive infinity \cite{Lohe:1979mh,Lohe:1980js}, {\it cf.} 
figure \ref{fig:sol6}. The potential for the harmonic fluctuations around this soliton is 
invariant under spatial reflection and the VPE can be straightforwardly computed using the 
methods described in chapter IV of Ref. \cite{Graham:2009zz}. As seen from the left panel in 
figure \ref{fig:vpe}, the VPE approaches negative infinity as $\alpha\to0$ 
\cite{Weigel:2016zbs,AlonsoIzquierdo:2011dy}. On the other hand, the classical energy maintains 
a finite positive value in that limit. In view of the above discussed $\lambda$ dependence 
this means that for any non-zero $\alpha$ there will always exist a $\lambda$ small enough 
such that the total energy is positive and the soliton is stable in the so-constructed 
parameter space.

The situation changes drastically for $\alpha=0$. Then a secondary vacuum emerges at 
$\varphi=0$. In that case two soliton solutions $\varphi_{\pm}$ exist that link $\varphi=0$ 
with $\varphi=1$ and $\varphi=-1$ with $\varphi=0$, respectively. Actually for tiny but non-zero 
$\alpha$, the soliton from above can be viewed as a combination of these two solutions whose 
separation increases as $\alpha\to0$. This may also be inferred from the profile functions
shown in figure \ref{fig:sol6}.
\bigskip

The potential for the harmonic fluctuations around the ($\alpha=0$)-soliton is not invariant under spatial
reflection. In particular the curvatures of the field potential around the minima at $\varphi=0$ and
$|\varphi|=1$ are different. As a result, a translational variance emerges \cite{Weigel:2017iup}: 
Let $x_0$ be the center of the soliton which can be determined either from the shift of argument of 
the soliton profile or from by the maximum of the corresponding classical energy density. Extended 
spectral methods reveal that the VPE changes by approximately $0.101$ per unit of $x_0$ 
\cite{Weigel:2016zbs}. (The sign of $\Delta E_{\rm vac}/\Delta x_0$  depends on which of the two 
solitons is considered. A similar variation has been observed as a force acting on the 
soliton \cite{Romanczukiewicz:2017hdu}.) The same variation is found for the VPE of background 
potential built from a barrier of width $x_0$ and a hight determined from the difference of the 
two curvatures. In contrast to $\alpha$, $x_0$ is a variational parameter that describes the shape 
of the soliton and for any given $\lambda$ we can choose $x_0$ such that the total energy is negative. 
Hence the ($\alpha=0$)-soliton is unstable at one-loop order.

The following picture emerges: As $\alpha\to0$ the soliton dwells in an ever increasing region of 
the ($\alpha=0$) secondary vacuum and the VPE decreases without bound. While for $\alpha\ne0$ that 
region may be large but still confined, for $\alpha=0$ it exceeds any boundary and the total energy 
becomes negative for any value of $\lambda$.

\section{THE SHIFMAN-VOLOSHIN SOLITON}

This soliton is based on the Bazeia model \cite{Bazeia:1995en} that contains two
real scalar fields governed by the Lagrangian
\begin{equation}
\lambda\mathcal{L}=\frac{1}{2}\left[\partial_\nu \phi\partial^\nu \phi
+\partial_\nu \chi\partial^\nu \chi\right]-U(\phi,\chi)
\qquad {\rm with}\qquad
U(\phi,\chi)=\frac{1}{2}\left[\phi^2-1+\frac{\mu}{2}\chi^2\right]^2
+\frac{\mu^2}{2}\phi^2\chi^2\,.
\label{eq:fpot} \end{equation}
Here the model parameter is $\mu\ge0$. For the exceptional value $\mu=2$ the model is 
equivalent to two independent $\phi^4$ models \cite{AlonsoIzquierdo:2012tw}.
The primary vacua are at $\chi=0$ and $\phi=\pm1$ while the secondary vacua are
given by $\chi=\pm\sqrt{\frac{2}{\mu}}$ and $\phi=0$. The curvatures, as expressed 
by the mass squared matrices, are related by ($\chi$ is particle one and $\phi$ is two)
\begin{equation}
M^2_{\rm sec.}=M^2_{\rm prim.}+\left(\mu-2\right)
\pmatrix{-\mu & 0 \cr 0 & 2}
\qquad {\rm with}\qquad M^2_{\rm prim.} =\pmatrix{\mu^2 & 0 \cr 0 & 4}\,.
\label{eq:curv}\end{equation}
The difference matrix has positive and negative entries (unless $\mu=2$), whence there is 
no predetermined ordering of the vacua.

\begin{figure}[ht]
  \centerline{\includegraphics[width=6.5cm,height=4cm]{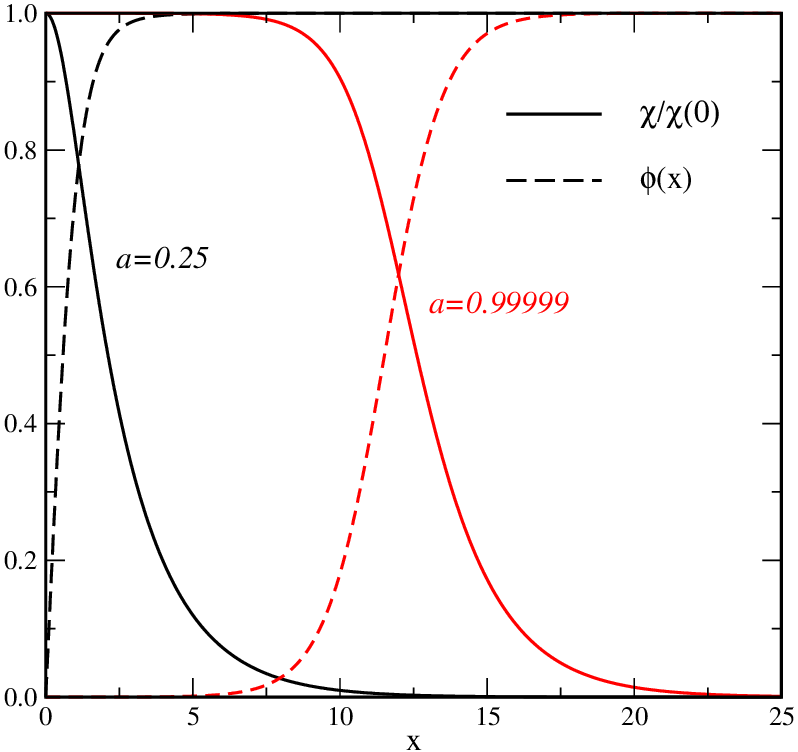}\hspace{1.9cm}
              \includegraphics[width=6.5cm,height=4cm]{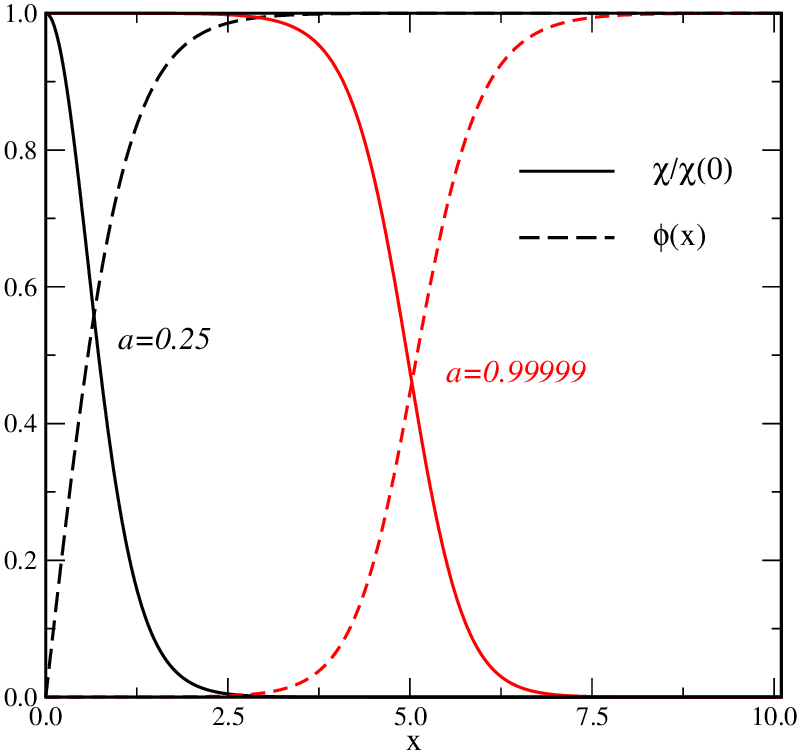}~}
  \caption{\label{fig:svsol}(Color online) The Shifman-Voloshin soliton with $x_0=0$ 
for $\mu=0.5$ (left panel) and $\mu=3.0$ (right panel) and different values of the 
variational parameter $a$. Figure adapted from Ref.~\cite{Weigel:2018jgq}.}
\end{figure}

The soliton profiles link the primary vacua and are characterized by two variational parameters: 
the center of the soliton, $x_0$ and the amplitude of the second field at the center, 
$\chi(x_0)=a\sqrt{\frac{2}{\mu}}$ with $|a|\le1$ \cite{Shifman:1997wg}. The classical
mass is degenerate with respect to these parameters. When $|a|\to1$ the soliton contains an 
ever increasing region of the secondary vacuum, as shown in figure \ref{fig:svsol}.
\bigskip

\begin{figure}[ht]
  \centerline{\includegraphics[width=7cm,height=4cm]{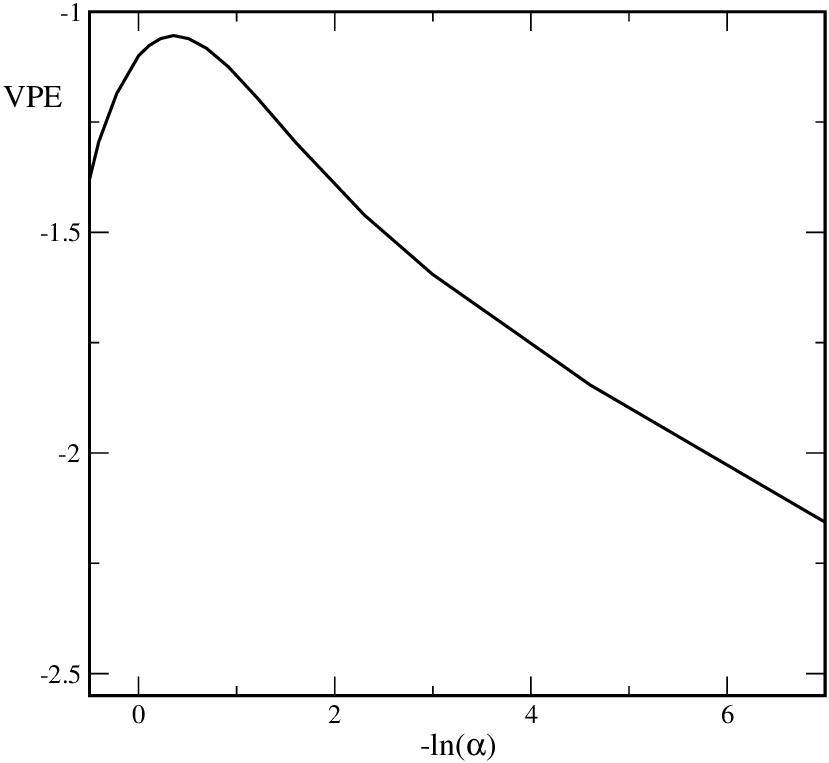}\hspace{1.3cm}
              \includegraphics[width=7cm,height=4cm]{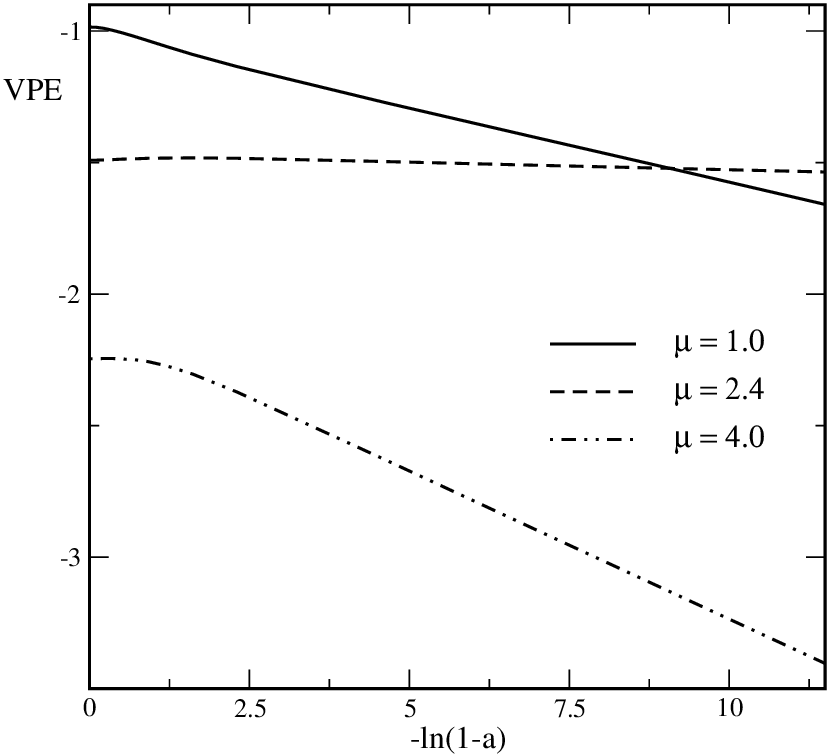}~~~}
  \caption{\label{fig:vpe}VPE of the $\phi^6$ model soliton as a function of the
model parameter $\alpha$ (left panel) and the Shifman-Voloshin soliton for various model
parameters $\mu$ as functions of the variational parameter $a$ (right panel). Figure adopted 
from data in Refs.~\cite{Weigel:2016zbs} and~\cite{Weigel:2018jgq}, respectively.} 
\end{figure}

Since the masses of the particles differ, a threshold emerges at the real momentum 
$k_{\rm th}=\sqrt{|\mu^2-4|}$. To compute the VPE this requires further extensions of the 
spectral methods that are described and applied in Refs. \cite{Weigel:2017kgy,Weigel:2018jgq}. 
Some of the numerical results of that analysis are shown in the right panel of figure \ref{fig:vpe}. 
Except for $\mu=2$ the VPE decreases logarithmically when $a$ approaches one. The reason for 
this exception simply is that $M^2_{\rm sec.}=M^2_{\rm prim.}$ because the model decouples 
into two $\phi^4$ kinks. Since $\mu=2.4$ is not too different from $\mu=2$, the slope of the 
corresponding line in figure~\ref{fig:vpe} is small but definitely negative.
\bigskip

In Ref. \cite{Weigel:2018jgq} it has been verified that, when $a$ is close to one, the VPE
of the Shifman-Voloshin soliton follows closely that of the background potential
$$
V(x)=\left[M_{\rm sec.}^2-M_{\rm prim.}^2\right]\Theta(L-|x|)
=\left(\mu-2\right) \pmatrix{-\mu & 0 \cr 0 & 2}\Theta(L-|x|)\,,
$$
where $L$ measures the size of the region in which the soliton assumes the secondary vacuum.
This potential mimics the transition of the soliton between the primary and secondary vacua.
Hence the existence of such a secondary vacuum and that it can be accessed by the soliton 
in an arbitrarily large regime cause the ever decreasing VPE.

Since $a$ is a variational parameter not related to the (fixed) model parameters $\lambda$ and
$\mu$, it will always be possible to construct a classical soliton such that the sum of the 
classical energy and the VPE is negative. Hence the classical soliton becomes unstable.

The conclusion on instability has been drawn from numerical results for the VPE in the
no-tadpole renormalization scheme. This conclusion remains valid for the physical on-shell
scheme \cite{Weigel:2018jgq}.

\section{CONCLUSION}
We have computed the vacuum polarization energies for classically stable solitons in 
$D=1+1$ models that have degenerate vacua with different curvatures in field space. A particular
feature of these solitons is that there are large regions in space where the soliton profiles 
assume, or at least arbitrarily closely approach, the secondary vacuum configuration. In those 
regions the VPE has a negative density and by increasing them, the total energy turns negative.
When this increase in governed by a variational rather than a model parameter, under which the
classical energy is bounded, the soliton is unstable, at least at one-loop order of the quantum 
corrections. We conjecture that such quantum instabilities occur whenever 
the soliton contains an unbounded region with the secondary vacuum.

\section{ACKNOWLEDGMENTS}
Parts of this presentation originate from a collaboration with N. Graham and M. Quandt whose
contributions are gratefully acknowledged.
This work is is supported in part by the National Research Foundation of South Africa (NRF)
by grant~109497.


\nocite{*}
\bibliographystyle{aipnum-cp}%

\end{document}